\documentclass[10pt,twocolumn,twoside]{IEEEtran}
\usepackage{cite}
\usepackage{graphicx}
\usepackage{epstopdf}
\usepackage[cmex10]{amsmath}
\usepackage[caption=false,font=footnotesize]{subfig}
\usepackage{stfloats}
\usepackage{multirow}
\usepackage{url}
\usepackage{amssymb}
\usepackage{amsmath}
\usepackage{balance}
\usepackage[linesnumbered,ruled]{algorithm2e}
\usepackage{diagbox}
\usepackage{booktabs}
\usepackage{float}
\usepackage{gensymb}
\newcommand{\tabincell}[2]{\begin{tabular}{@{}#1@{}}#2\end{tabular}}

\usepackage{xcolor}

\begin{document}

\title{True-data Testbed for 5G/B5G Intelligent Network}

\author{Yongming~Huang, Shengheng~Liu, Cheng~Zhang, Xiaohu~You, and~Hequan~Wu

\thanks{This work was supported in part by the National Key R\&D Program of China under Grant No. 2018YFB1800801, and the National Natural Science Foundation of China under Grant Nos. 61720106003 and 62001103. \emph{(Corresponding author: Xiaohu~You.)}}
\thanks{Yongming~Huang, Shengheng~Liu, Cheng~Zhang, and Xiaohu~You are with the National Mobile Communications Research Laboratory, Southeast University, Nanjing 210096, China. E-mail: \{huangym; s.liu; zhangcheng\_seu; xhyu\}@seu.edu.cn.}
\thanks{Hequan~Wu is with the China Information and Communication Technology Group Corporation, Beijing 100083, China. E-mail: wuhq@cae.cn.}
\thanks{The authors are also with the Purple Mountain Laboratories, Nanjing 211111, China.}

}

\maketitle

\begin{abstract}
Future beyond fifth-generation (B5G) and sixth-generation (6G) mobile communications will shift from facilitating interpersonal communications to supporting Internet of Everything (IoE), where intelligent communications with full integration of big data and artificial intelligence (AI) will play an important role in improving network efficiency and providing high-quality service. As a rapid evolving paradigm, the AI-empowered mobile communications demand large amounts of data acquired from real network environment for systematic test and verification. Hence, we build the world's first true-data testbed for 5G/B5G intelligent network (TTIN), which comprises 5G/B5G on-site experimental networks, data acquisition \& data warehouse, and AI engine \& network optimization. In the TTIN, {true} network data acquisition, storage, standardization, and analysis are available, which enable system-level online verification of B5G/6G-orientated key technologies and support data-driven network optimization through the closed-loop control mechanism. This paper elaborates on the system architecture and module design of TTIN. Detailed technical specifications and some of the established use cases are also showcased.
\end{abstract}


\bigskip

\section{Introduction}
\label{sec:intro}

\IEEEPARstart{A}{t} present, the broad deployment and commercial operation of fifth-generation (5G) mobile communication are speeding up. Meanwhile, the blue prints of beyond 5G (B5G) and six-generation (6G) mobile communication have already been discussed and explored since 2018, where numerous emerging technologies ranging from the terahertz communication, satellite-territorial-integrated networks to artificial intelligence (AI) have been envisioned as the potential key enablers \cite{Ref0, Ref1, Ref2}. Compared with state-of-the-art 5G, future B5G and 6G are not only expected to provide wider frequency band, higher transmission rate, shorter delay, and wider coverage, but also remarkably higher network intelligence \cite{Ref3}. It is commonly accepted that, data-driven AI innovations will play a vital role in enabling the B5G and 6G to be much more intelligent in the self-learning, self-optimizing, and self-managing capabilities \cite{Ref4}.

The past few years have witnessed numerous data-driven innovations in network automation, optimization, and management, which dramatically improve the level of intelligence of the wireless communication networks. Nearly every facet of the networks including intelligent beam management \cite{Ref5, Ref6, Ref7, Zhang20a, Zhang20b, Ref10}, mobility management \cite{Ref8, Ref9}, interference coordination \cite{Ref11, Ref12}, power controlling \cite{Ref13, Ref14, Ref15}, energy conservation \cite{Ref16, Ref17}, network slicing \cite{Ref18, Ref19}, network traffic management \cite{Ref21, Ref22}, edge computing \cite{Ref23, Ref24, Ref25, Ref26}, and network anomaly detection \cite{Ref27} are being reshaped. Rich data is among the indispensable elements that guarantee the effectiveness of AI engines \cite{Ref28}. However, as neither true-data testbed nor real (or even verisimilar) network data is publicly accessible yet, AI-empowered 5G/B5G research is confined to numerical simulations with simplified assumptions and synthetic datasets. This in turn dramatically deteriorates the practical effectiveness of data-driven innovations. The gross inadequacies in the true data and data source have become the bottleneck that hinders the evolution of the intelligent wireless communication networks. As such, there is an urgent need for establishment of an open, flexible, scalable, and intelligent 5G/B5G testbed to fill the gaps and support the on-line optimization for AI algorithms. We envision that, with the assistance of the advanced data sensing, aggregation, storage, and analysis technologies, the potential of rich network data acquired from the true-data testbed can be fully unleashed to enable intelligent networks.

\begin{table*}[!hb]
	\renewcommand\arraystretch{1.2}
	\centering
	\caption{Comparisons of the proposed TTIN with the state-of-the-art 5G experimental networks.}
	\label{tab:1}\vspace{2em}
	\begin{tabular}{|p{58pt}||p{36pt}|p{36pt}|p{38pt}|p{102pt}|p{114pt}|}\hline
		Experimental Network & SDN\newline Controlling & Data\newline Warehouse & AI Engine & Supported Applications & Network Segment\\\hline\hline
		TTIN   & Supported & Supported & Supported & UAV, IoV, video service, smart grid, AR/VR & Radio access network, core\newline network, transport networks\\\hline
		5G-EmPOWER \cite{Ref29} & Supported & Not\newline mentioned & Not\newline mentioned &	RAN slicing, load-balancing, mobility management & Radio access network \\\hline
		CTTC \cite{Ref30} & Supported & Not\newline mentioned & Not\newline mentioned & Mobile broadband services, IoT service & Wireless access \& backhaul\newline networks, metro aggregation network, core transport network\\\hline
		ADRENALINE \cite{Ref31} & Supported & Not\newline mentioned & Not\newline mentioned & IoT service, end-to-end 5G & Optical transport network \\\hline
		5G-DRIVE \cite{Ref35}  & Supported & Not\newline mentioned & Not\newline mentioned & Enhanced mobile broadband, vehicle-to-everything & Wireless and fibre optic backhaul network, transport network\\\hline
	\end{tabular}
\end{table*}

In fact, a few pioneering attempts have been made along this pathway in the recent years. As summarized in Table~\ref{tab:1}, EmPOWER \cite{Ref29} and ADRENALINE project are among the most inspiring work along the pathway, which are respectively built for the 5G radio access network (RAN) are 5G optical transport network (OTN) prototype verifications. More recently, the CTTC \cite{Ref30} has developed an end-to-end 5G platform to test the Internet of thing (IoT) and mobile broadband services by integrating several existing testbeds such as ADRENALINE \cite{Ref31}, GEDOMIS \cite{Ref32}, EXTREME \cite{Ref33}, CASTLE and IoTWorld \cite{Ref34}. In 2018, a research collaboration between China and European Union on 5G networks has been initiated. This project, known as 5G-DRIVE, planned to unite diverse testbeds distributed in China, United kingdom, Finland and Italy to trial and validate key technologies of 5G networks operating at $3.5\;{\rm{GHz}}$ bands for enhanced mobile broadband (eMBB) and $3.5/5.9\;{\rm{GHz}}$ bands for vehicle-to-everything (V2X) scenarios, and this collaboration is currently still moving forward \cite{Ref35}. With the aid of the data warehouse and the AI engine, the potential of rich network data acquired can be fully exploited to support diverse AI-empowered network optimization applications and to enable intelligent networks for B5G/6G, which have not been reported in the existing networks. At this point, to build a demonstrative fully functional 5G/B5G platform that includes 5G radio access, metro aggregation and transport networks with heterogeneous wireless/optical technologies is quite challenging. Numerous interfaces need to be opened and comprehensive network data from different layers should be collected, pre-processed, stored, and analyzed efficiently to support the intelligent closed-loop control and the on-line AI optimization in diverse use cases.

\begin{figure*}[!hb]
\centering
\includegraphics[width=0.75\textwidth]{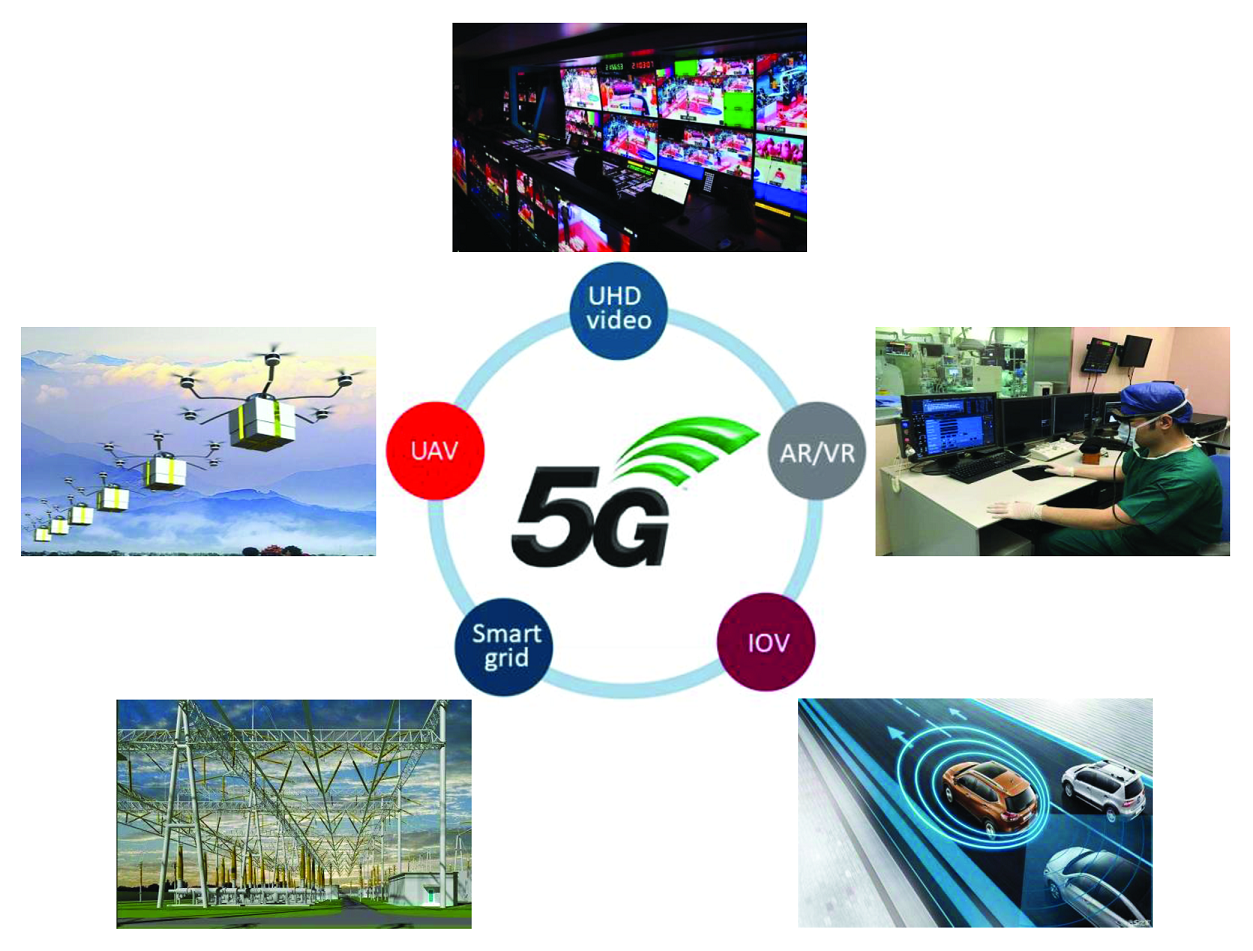}
\caption{Typical 5G applications.}
\label{Fig.1}
\end{figure*}

In this context, the world's first true-data testbed for 5G/B5G intelligent networks (TTIN) is released. The TTIN covers core networks, optical transport networks, and 5G radio access networks with millimeter-wave (mmWave) base stations deployed. The 5G/B5G intelligent network is built in the standalone (SA) mode complying with the 3GPP standards and aims to support a broad spectrum of B5G/6G-oriented researches ranging from smart grid, smart city, smart health care, and many more. The start-up coverage range of the 5G/B5G network is confined to the China Wireless Valley (CWV), Jiangning District, Nanjing, China. In this 5G/B5G intelligent network, open interfaces, on-line AI optimization and the closed-loop control are available, where complete network data can be collected from the terminals, base stations, core networks and transport networks in real time and further sent into the big-data platform for standardized pre-processing procedure including data cleaning, labeling, classifying, desensitization, and storage. The pre-processed data is then analyzed in the intelligent computing platform (ICP) using various AI algorithms. {The AI algorithms are trained and optimized on line} and the intelligent control orders are finally generated which provide feedbacks and control the network modules in a closed-loop manner. Reap the rewards of the closed-loop control mechanism, the hierarchical open-radio access network (O-RAN) management structure, and the advanced network optimization approaches such as intelligent beam management and interference coordination, the network performance of the TTIN in terms of data rate, latency, and transmission capacity has been significantly enhanced. In addition, TTIN will continue to grow as new spectrum resources (e.g., mmWave \cite{Ref38}, terahertz, and visible light), extended coverage (i.e., ground, air, and space), and new AI-powered network management ability are introduced. Compared with the state-of-the-art 5G platforms, TTIN can potentially better serve the typical 5G applications such as video service, UAV communications, and Internet of vehicles (IoV) as shown in Fig.~\ref{Fig.1}.

Following this introductory section, the overall architecture of the system, the description of each key module, and the supported use cases will be successively presented in the next three sections. This paper is concluded in Section V.

\section{System Architecture}

\begin{figure*}[!hb]
\centering
\includegraphics[width=1\textwidth]{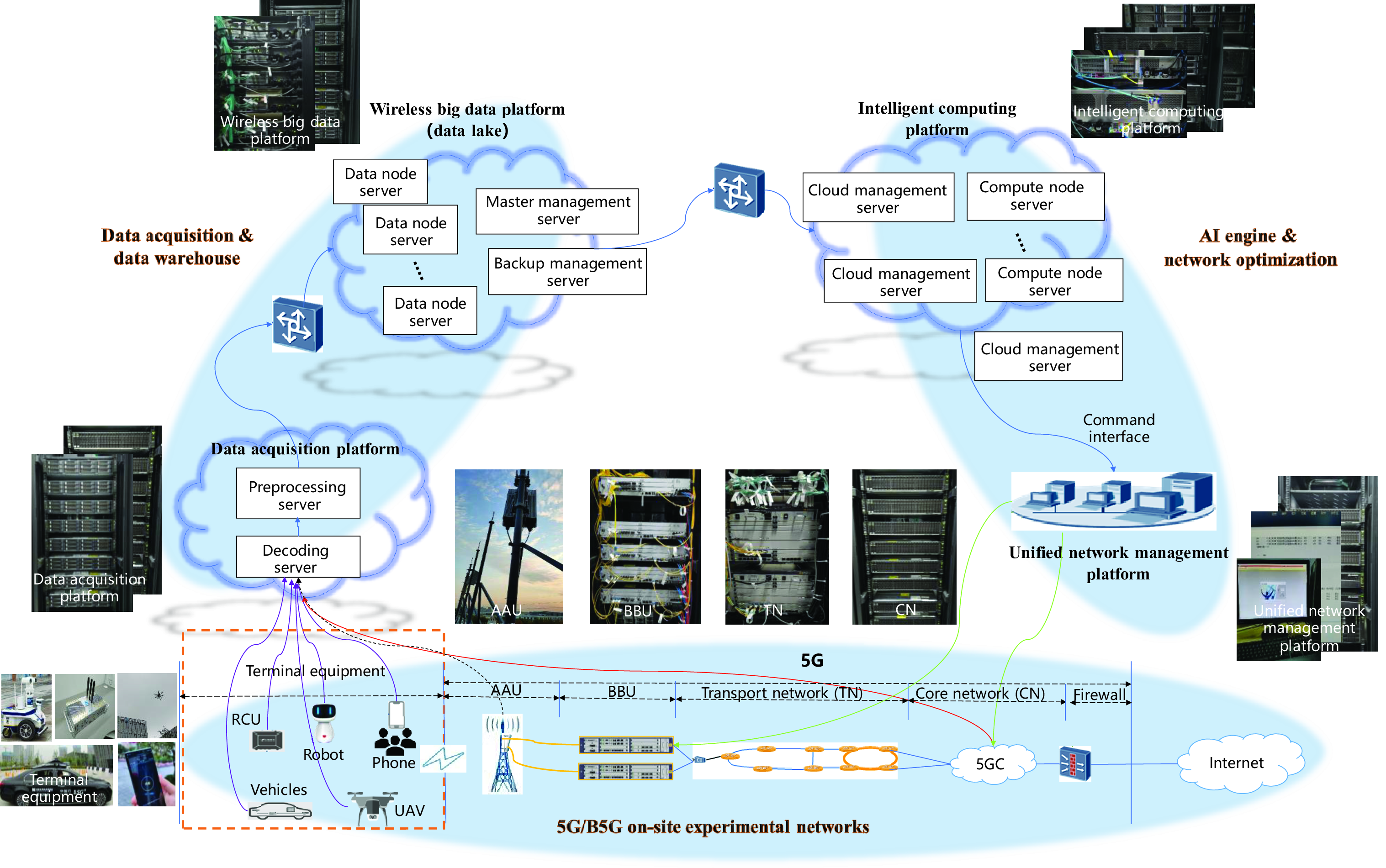}
\caption{System architecture of TTIN.}
\label{Fig.2}
\end{figure*}

\subsection{Architecture}

To allow true-data experiment of schemes and methodologies for intelligent mobile networks, we have built the world's first true-data testbed for real-time big data acquisition, storage, analysis, and intelligent closed-loop control. As shown in Fig.~\ref{Fig.2}, the TTIN is composed of 5G/B5G {on-site experimental} networks, data acquisition \& data warehouse, and AI engine \& network optimization. In the TTIN, commercial off-the-shelf equipment and devices have been deployed to establish an analogue of the real network environment. The key devices in the 5G/B5G on-site experimental networks consist of the Huawei’s AAU5613, the Huawei’s BBU5900, NE20E-S routers, Huawei’s servers, disk arrays, and optical transceivers. Moreover, commercial servers and the industrial Hadoop platform are adopted in the data acquisition platform and the wireless big data platform respectively. In the intelligent computing platform, Xeon servers with Tesla V100 GPUs and NVIDIA T4 GPUs serve as the powerful computer cluster. To supply with the advanced network management capability, the Huawei’s U2020 network management system is utilized in the unified network management platform. In addition, different terminal equipment including the Dingli pilot RCU, Huawei Mate30 mobile phones, DH X1100 UAVs, commercial robots and vehicles are also adopted to test and verify the performances of diverse 5G/B5G applications for the TTIN. Each key module is briefly introduced in the following and more details will be given in Section~\ref{sec:3}.

\subsubsection{5G/B5G on-site experimental networks}
5G/B5G experimental network adopts 3GPP R15 SA architecture, which contains macro base stations, active antenna units (AAUs), small stations, base band units (BBUs), a complete set of core network, transmission networks, and a network management system. The working frequency ranges from $3.5\;{\rm{GHz}}$ to $3.6\;{\rm{GHz}}$. TTIN can be used as a commercial network to support diverse 5G services, in which the data interfaces of core network and transmission network are all opened to provide comprehensive network state data in real time.

\subsubsection{Data acquisition \& data warehouse}
Heterogeneous network data from multiple sources including the terminals, base stations, core networks, and road drive tests in the 5G/B5G networks are collected via the data access modules. They are then aggregated into the highly-integrated dataset. After the standard data pre-processing, which includes the data cleaning, classification, association, construction, and storage operations, the aggregated network data is sent to the data storage/computing module, where the pre-processed data is further processed using the business mining modelling module to yield organized data according to the specific subjects. The basic warehouse is maintained for different sources and, on this basis, correlation analysis, endogenous factor (EF) extraction, and knowledge graph (KG) formation are implemented to establish subject/service-oriented data warehouse, which represents a most important technical highlight of this TTIN platform. The stored data in the data warehouse are time-varying, which are updated and replenished by the fresh network data sensed from the dynamic network environment. Nevertheless, the data warehouse is stable over a modest time period since query analysis dominates and every few data updating or removal is performed. This subject-oriented, highly-integrated, time-varying, and stable 5G data warehouse can supply organized datasets for diverse intelligent network optimization applications, which provides a unique, first-of-its-kind and the only repository for investigators from all over the globe.

\subsubsection{AI engine \& network optimization}
The AI engine and network optimization are enabled by the ICP and the UNMP. The AI engine serves as a general hardware $\&$ software infrastructure for intelligent computing, which consists of four training node servers, two inference node servers, one set of basic software for AI cloud platform equipped with NVIDIA V100 and Tesla T4 GPUs. Additionally, the UNMP is utilized to realize centralized network operation and maintenance functions for the TTIN, which manages different network elements deployed in the access radio networks, core networks, and optical transmission networks. Supported by the AI engine and the UNMP, the TTIN is capable of controlling numerous adjustable parameters of corresponding network elements to realize intelligent network optimization.

\subsection{Operation mode and characteristics}
TTIN works in a closed-loop operation mode, which starts from data acquisition from the 5G/B5G networks in implementing different network optimization tasks. With the support of sufficient data, researchers are able to design and train optimization models with the support of wireless  big data platform and ICP. Then, the parameters are sent to the UNMP, and the network is finally adjusted accordingly. The above procedures form an intelligent communication closed-loop optimization system including data acquisition, model design, intelligent optimization, model deployment, and network verification.

In summary, TTIN has the following characteristics:

\begin{itemize}
\item Advanced architectures and comprehensive functions. The architecture of the TTIN follows the 3GPP R15 standards, which covers radio access networks, core networks, and optical transmission networks. The TTIN meets commercial operation standards and is capable of effectively supporting different B5G/6G-oriented innovations and testing diverse use cases including the vehicular network, industrial internet, and other researches from the chip level to the system level.

\item Open interfaces and intelligent closed-loop control mechanisms. The data interfaces in TTIN are opened to support the comprehensive data collection function of the DAP. The collected network data is further sent into subsequent operation modules including the data cleaning, data labeling/storage in the wireless big data platform (WBDP), data analysis/computation in the ICP, order releasing/feedback monitoring in the UNMP, which constitutes an intelligent closed loop for the network management.

\item Open-source ecosystems and ever-evolving capabilities. TTIN aims to establish an open community to test and verify different B5G/6G-oriented innovation solutions. We envision that many important network equipment and base stations will be gradually opened and become white-boxs. Meanwhile, new spectrum resources, extended coverage, and AI-powered network management ability will promote the evolvement of the TTIN.
\end{itemize}

\section{Key Modules}
\label{sec:3}

We elaborate the key modules of TTIN in this section.

\subsection{5G/B5G on-site experimental networks}

\subsubsection{5G radio access network}

\begin{figure*}[!hb]
\centering
\includegraphics[width=0.92\textwidth]{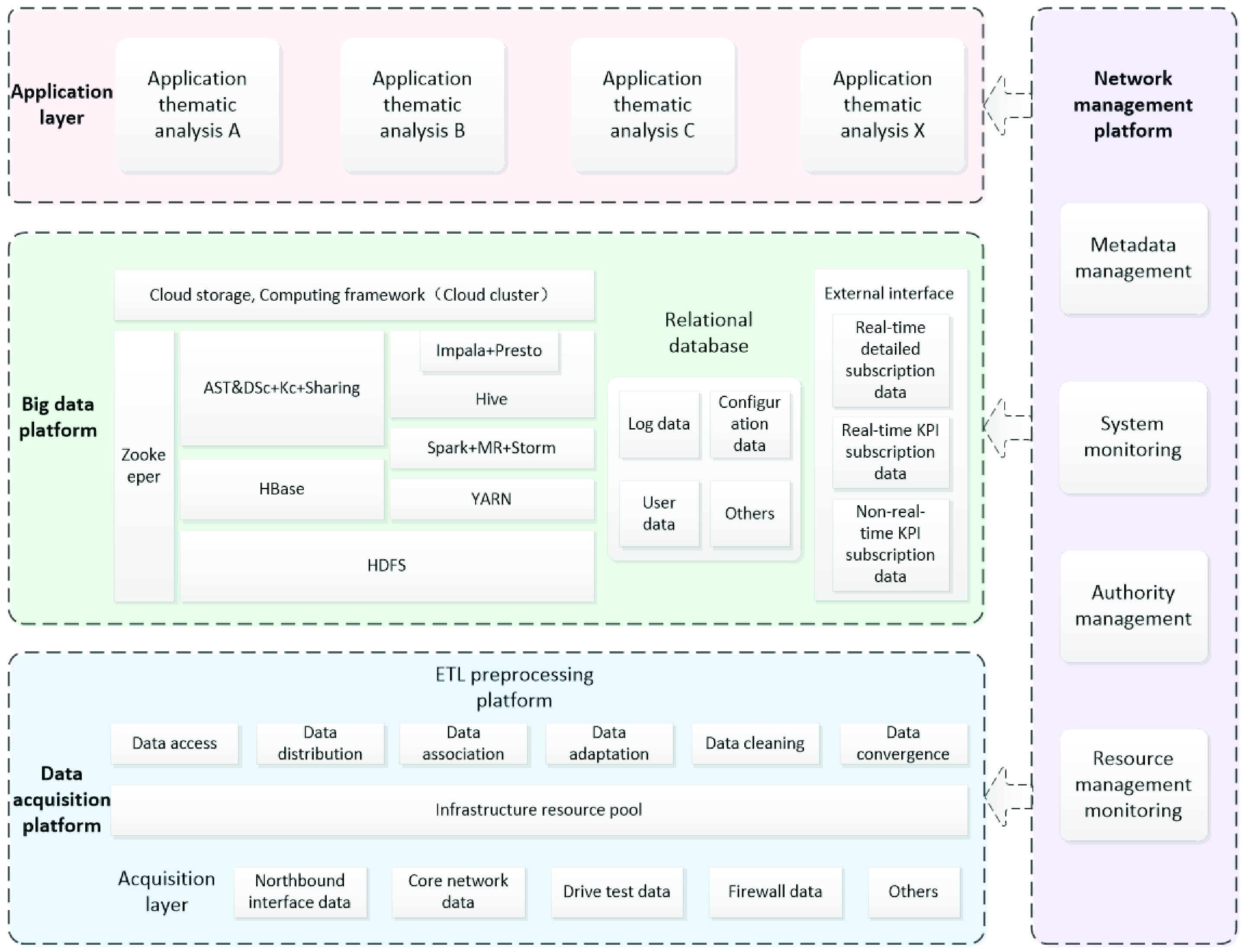}
\caption{Framework of the intelligent wireless communication platform.}
\label{Fig.3}
\end{figure*}

In this 5G/B5G on-site experimental network, the AAU and the BBU serve as the key network elements for the 5G radio access network. The AAU is specially designed to reduce the resource consumption of the 5G base stations, where the functions of the remote radio unit (RRU) and the massive multiple-input multiple-output (MIMO) antenna are combined. Benefited from the massive MIMO antenna array in the AAU, the transmission capacity and quality for end users have been dramatically improved, where $200\;{\rm{MHz}}$ maximum bandwidth is available to support different spectrum resource utilizations in C-Band and much higher diversity gain is obtained. Aside from the large bandwidth, to support the flexible beam management, the vertical and horizonal angles of the antenna are also adjustable for enabling the precise beam management to provide dynamic coverage and enhance the quality of transmission. Here, AAUs supporting 64T/64R and macro base stations are distributed outside the building. Meanwhile, small stations and pHUBs are deployed inside the buildings to cover eight floors. On the other hand, the main functions of the BBU are baseband signal processing, base station system management (operation maintaining, order processing, and system clock management), physical interface provisioning, operation and maintenance center (OMC) channel provisioning, and environment information exchanging. The specific configurations of the 5G radio access network are listed in Table~\ref{tab:2}.

\begin{table}[!htpb]
	\renewcommand\arraystretch{1.2}
	\centering
	\caption{The configurations of the 5G radio access network}
	\label{tab:2}\vspace{2em}
	\begin{tabular}{|p{166pt}|l|}\hline
		Device type&Description\\\hline
		Massive MIMO antenna&\tabincell{l}{64T64R}\\\hline
		\tabincell{l}{AAU}&\tabincell{l}{12}\\\hline
		\tabincell{l}{BBU}&\tabincell{l}{5}\\\hline
		\tabincell{l}{pHUB}&\tabincell{l}{8}\\\hline
		\tabincell{l}{Macro base station}&\tabincell{l}{9}\\\hline
		\tabincell{l}{Small base station}&\tabincell{l}{50}\\\hline
	\end{tabular}
\end{table}

\subsubsection{5G core network and transport network}

In the backhaul link, the traffic flows from the BBUs are aggregated into the transport networks, which consist of routers and optical transceivers. In the routers, the distributed processing architecture is adopted, where the controlling, switching, and relaying functions are separated, such that the control and user planes are guaranteed to be independent. The optical transport network is used to realize high-speed links over a distance of several kilometers in the CWV. Moreover, for the 5G core networks, 22 servers, two disk arrays and seven switches are deployed to enable the service-oriented core networks in the SA architecture, which supports diverse 5G services, i.e. the gigabit mobile broadband, 4K/8K video, AR/VR, factory automation, automatic driving and intelligent city, and different access ways including the 2/3/4/5G, Wi-Fi, narrow band Internet of Things (NB-IoT), fixed access, and unlicensed spectrum. In the core network, the network slicing techniques and the service-oriented architecture are utilized to provide different quality of service, a simplified network, and unified network operations. Additionally, the 5G/B5G experimental network complies with 3GPP R15 standalone architecture, where new base stations, backhaul links, and core networks are established, which are independent with the network infrastructure legacy in 2G/3G/4G. Furthermore, the network performances in terms of the uplink/downlink peak rates, uplink/downlink user experience data rates, end-to-end delay, the success rate of handover and the traffic density are comprehensively tested by the Dingli Corporation, where the files with the size of $50\;{\rm{GB}}$ and $20\;{\rm{GB}}$ are used to download and upload through the FTP, and the network performances of 17 cells distributed in CWV serve as the testing samples. The key performance parameters of the established 5G experimental networks are concluded in Table~\ref{tab:3}.

\begin{table}[!htpb]
	\renewcommand\arraystretch{1.2}
	\centering
	\caption{Key performance parameters of the 5G experimental network}
	\label{tab:3}\vspace{2em}
	\begin{tabular}{|p{142pt}|l|}\hline
	Performance parameter&Description\\\hline
		Uplink peak rate/user&\tabincell{l}{$261\;{\rm{Mbps}}$}\\\hline
		\tabincell{l}{Downlink peak rate/user}&\tabincell{l}{$1.18\;{\rm{Gbps}}$}\\\hline
		\tabincell{l}{Uplink user experience data rate}&\tabincell{l}{$170\;{\rm{Mbps}}$}\\\hline
		\tabincell{l}{Downlink user experience data rate}&\tabincell{l}{$940\;{\rm{Mbps}}$}\\\hline
		\tabincell{l}{End-to-end delay}&\tabincell{l}{$14\;{\rm{ms}}$}\\\hline
		\tabincell{l}{Success rate of handover}&\tabincell{l}{100\%}\\\hline
		\tabincell{l}{Traffic density}&\tabincell{l}{$1.045\;{\rm{Tbps/km^2}}$}\\\hline
	\end{tabular}
\end{table}

\subsubsection{Supported terminal types}
Combined with advanced 5G network elements and the SA architecture, the TTIN supports various terminals including mobile phones, vehicles, UAVs, high-definition cameras, drive test equipment, and customer premise equipment (CPE).

\subsection{Data acquisition \& data warehouse}

\begin{figure*}[!htpb]
\centering
\includegraphics[width=0.9\textwidth]{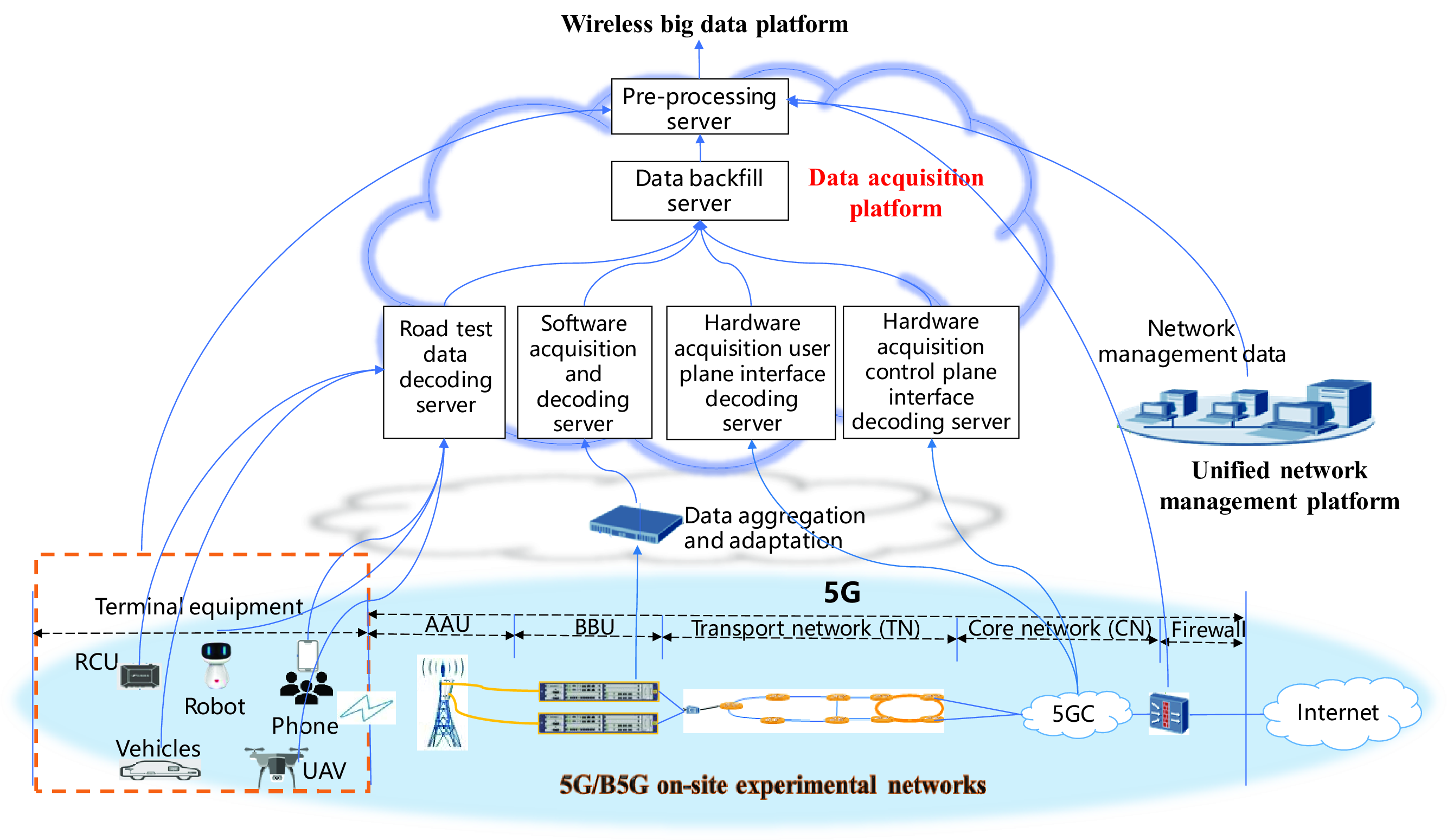}
\caption{Framework of the data acquisition platform.}
\label{Fig.4}
\end{figure*}

\subsubsection{Data acquisition platform}
As shown in Fig.~\ref{Fig.4}, DAP is responsible of collecting and pre-processing comprehensive network data from different sources.

\begin{table}[!htpb]
	\renewcommand\arraystretch{1.2}
	\centering
	\caption{Typical network data collected in the DAP}
	\label{tab:4}\vspace{2em}
	\begin{tabular}{|p{70pt}|l|}\hline
		Data types&Description\\\hline
		Air interface data&\tabincell{l}{Data from PHY/ MAC/ RLC/ PDCP/\\ RRC/ NAS layers}\\\hline
		\tabincell{l}{Software\\ acquisition data}&\tabincell{l}{Raw data of Uu/Xn/X2/E1/F1 and\\ other interfaces from the BBU}\\\hline
		\tabincell{l}{Core network\\ control plane data}&\tabincell{l}{Interface data of N1/ N2/ N4/ N5/ N6/\\ N7/N8/ N9/ N10/ N11/ N12/ N13/\\ N14/ N15/N16/ N17/ N18/ N19/ N20/\\ N21/ N22/ N23/ N24/ N26/ N27/ N28/\\ N29/ N30/ N31/ N35/ N36/ N40}\\\hline
		\tabincell{l}{Core network user\\plane data}&\tabincell{l}{Protocol data of N3\_S1U/ N3\_MMS/ \\N3\_HTTP/ N3\_HTTPS/ N3\_DNS/\\ N3\_FTP/ N3\_EMAIL/ N3\_SIP/\\ N3\_RTSP/ N3\_COAP}\\\hline
		\tabincell{l}{Pilot Matrix DT data}&\tabincell{l}{Signaling data from L1/ L2/ L3 layers}\\\hline
		\tabincell{l}{Network\\ management data}&\tabincell{l}{Network management data derived\\ from the northbound interface,\\ including performance, alarm,\\ configuration and other indicators}\\\hline
		Firewall data&\tabincell{l}{Logs and other data}\\\hline
	\end{tabular}
\end{table}

{\textbf{Data acquisition}}:
For data acquisition, various types of data including the air interface data, core network user interface data, core network control plane data, northbound interface data, and field test data as shown in Table~\ref{tab:4}, are accessed to the DAP through the safe data transfer protocol (SDTP) and secure file transfer protocol (SFTP). SDTP is adopted for real-time streaming data transmission. SFTP is used for the external data representation  data transmission under the cases with low real-time requirements. The transmitted data can be encoded in form of CSV or TXT.

{\textbf{Data preprocessing}}:
For data preprocessing, the reading, transformation, cleaning and encryption for streaming data and text data from the DAP are available. The major procedure is composed of the data cleaning, data transformation and data loading. In the data cleaning, irregular data and inconsistent data is removed for eliminating the inconsistency of data. In the data transformation, the transformation of unified data coding, data type and data format are realized. In the data loading, cleaned data is loaded into the Hadoop for the data access.

\subsubsection{{Wireless big data platform}}

In the WBDP, the platform deployment and construction are based on the open-source Hadoop platform widely used in the industry. Here, the Hadoop platform is utilized for the data warehousing, data storage and data calculation, which can provide visual interfaces for data analysis and query. Moreover, common functions such as security management and audit, operation and maintenance management and performance monitoring are also available. The major platform components include Hadoop distributed file system (HDFS) distributed file system, HBase column storage database, Hive data warehouse, Impala query, Presto query and yet another resource negotiator (YARN) resource management. In addition, combined with software-hardware integration and practical cases development experience, following enhanced functions are provided for the WBDP:

{\textbf{Hadoop Tuning}}: Besides basic functions provided above, this WBDP is capable of realizing the fine tuning and optimizations including the hardware optimization, operating system optimization, Hadoop configuration optimization and application optimization according to specific requirements.

{\textbf{Heterogeneous Storage}}:
For different types of network data, the WBDP provides the heterogeneous storage service. Cold data is migrated into the cheap storage (archive) while part of the hot data and the intermediate data generated by caching I/O intensive tasks are sent into the SSD, which improves the query and calculation speed and saves the cluster storage resources.

{\textbf{Stream Data Warehousing}}:
A streaming data warehousing platform is developed to realize task definition in the form of SQL, warehousing program parsing SQL, installing configuration rules for data analysis and storage of stream data statistics. According to the resource list of nodes in the WBDP, the collected network data is evenly distributed to corresponding node and abundant log information is provided to record the error log separately which is convenient for analyzing and correcting problems Positive measures.

{\textbf{HBase Optimization}}:
Object-oriented design and development method are adopted to encapsulate the operation of HBase. The complexity of HBase is transparent to developers, which is conducive to improve development efficiency and reduce bugs in development.

\subsection{AI engine \& network optimization}

\subsubsection{Intelligent computing platform}

A general intelligent computing infrastructure integrated software and hardware is provided in ICP. With the support of advanced network data sensing, gathering and preprocessing, different B5G/6G-oriented AI-empowered innovation solutions can be developed and deployed through this ICP. For instance, numerous radio resources of the TTIN can be intelligently allocated according to service requirements and the collected channel data, which generates dynamic allocating orders and further controls different network elements through the UNMP for the effectiveness verification. The specific hardware facilities and the AI cloud platform are illustrated below:

{\textbf{Hardware facilities}}:
The hardware facilities of ICP consist of seven high-performance Xeon servers with each have more than 25 CPUs. Among the seven servers, one server is used as the management node, four servers serve as training nodes and two servers are regarded as inference nodes. 12 Tesla V100 $32\;{\rm{GB}}$ GPUs are adopted as training node servers and 16 NVIDIA $16\;{\rm{GB}}$ T4 GPUs are used as inference node servers. With the supports of high-performance GPUs,  ICP is capable of implementing large-scale machine learning and deep learning computing tasks.

{\textbf{AI cloud platform}}:
AI cloud platform plays an important role in providing fundamental application services including the basic deep network framework and typical pre-training models. The supported languages and frameworks are listed in Table~\ref{tab:5}.

\subsubsection{{Unified network management platform}}

In the network management platform, the Huawei's U2020 network management system (U2020 for short) is adopted to manage different mobile network equipment centrally. The management system utilizes an open architecture and various types of network elements are connected through the network element adaptation layer, providing basic network management functions and rich optional functions including the configuration management, performance management, fault management, security management, log management, topology management, software management and system management. The server software of the network management platform is composed of the main version software and the adaptation layer software. The main version software realizes the system functions and the adaptation layer software completes the adaptation of different network element interfaces. Meanwhile, the management for new network elements can be realized by adding the corresponding adaptation layer software. The openness of this network management platform enables it to support the management for Global System for Mobile Communications (GSM), Universal Mobile Telecommunications System (UMTS), LTE, 5G and other wireless networks, WLAN, eRelay, SingleDAS, core network and next generation network (NGN). When the network is evolved, the network management platform can be upgraded at the same time.

In addition, this network management platform also supports other advanced functions including the network health check, remote network element upgrade, automatic base station planning, neighbor cell automatic optimization, remote base station debugging and testing, equipment panel, engineering alarm setting, and RAN sharing management.

\section{Supported Use Cases}

By harvesting the comprehensive data sensing, standard data pre-processing, and powerful AI analysis, the TTIN facilitates the intelligent closed-loop network optimization, where comprehensive network data can be gathered from the air interface, the core network user interface, core network control plane, and the northbound interface and then sent into the big data platform for standardized pre-processing procedure including data cleaning, data classification, data association, data construction and data storage according to different themes of applications. The pre-processed data is further analyzed in the ICP through diverse AI models and the intelligent control orders are finally emerged which feedback control the different network elements distributed in the network in the closed-loop way. In the TTIN, time-varying network states are sensed efficiently and flexible optimization operations can be utilized to improve network performances and guarantee high quality of service. Therefore, the TTIN is capable of supporting diverse network optimization tasks ranging from the RAN planing and deployment, massive MIMO, interference coordination, mobility management, fraud detection, load balancing, network slicing, MEC to the QoE optimization to realize larger transmission capacity, wider signal coverage, shorter network delay and higher network reliability. There are several examples of those supported use cases in the TTIN elaborated as follows:

\begin{table}[!hb]
	\renewcommand\arraystretch{1.2}
	\centering
	\caption{Supported functions of AI cloud platform}
	\label{tab:5}\vspace{2em}
	\begin{tabular}{|p{86pt}|l|}\hline
		Supported functions&Description\\\hline
		Supported languages&\tabincell{l}{C/C++/Java/Python...}\\\hline
		\tabincell{l}{Integrated development \\environments (IDEs)}&\tabincell{l}{VS/Pycharm/Anaconda...}\\\hline
		\tabincell{l}{Deep learning \\ frameworks}&\tabincell{l}{Tensorflow/Pytorch/\\ PaddlePaddle/Mxnet...}\\\hline
		\tabincell{l}{Pre-trained models}&\tabincell{l}{VGG 16, VGG 19,\\ Inception V1$-$V4, \\ResNet 50$-$101$-$152,\\ MobileNet V1$-$V2,\\ BERT, Transformer, GPT$-$2 ...}\\\hline
	\end{tabular}
\end{table}

\subsection{Intelligent Throughput Optimization}

Throughput is one of the key network performance parameters need to be optimized for the enhancement of network capabilities. Based on cognitive radio theory, researchers is capable of designing the power allocation and system parameter configuration elaborately to improve the throughput. In the conventional optimization methods, throughput optimization can be only implemented through limited simulations. With the support of TTIN, the researchers are allowed to test the develop different intelligent optimization innovations through experimental networks and practical network data.

\begin{figure}[!htpb]
\centering
\includegraphics[width=0.48\textwidth]{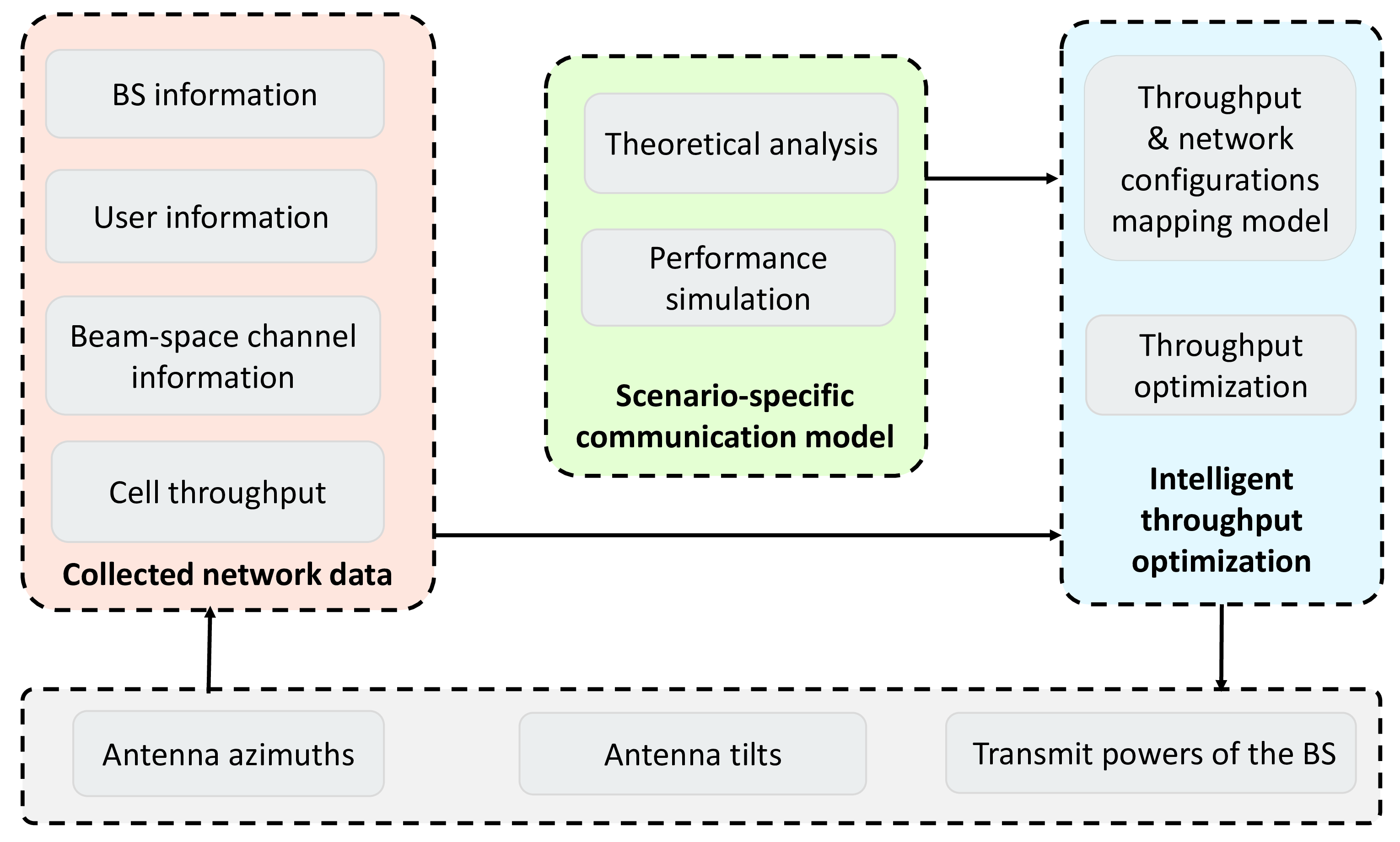}
\caption{Schematic diagram of the intelligent throughput optimization approach.}
\label{Fig.5}
\end{figure}

The schematic diagram of the intelligent throughput optimization approach is displayed in Fig.~\ref{Fig.5}. In order to realize high efficiency and high reliability, the approach driven by both network data and the theory model is proposed. On the one hand, as given in Table~\ref{tab:6}, comprehensive network state data can be collected from the TTIN, including the BS information (the BS positions, the transmit powers, the antenna azimuths, and antenna tilts), user information (the user positions and the user data rates), beam-space channel information, i.e., the reference signal received powers (RSRP) and signal to interference and noise ratios (SINR) of directional reference signals, such as synchronization signal block (SSB) and channel state information reference signal (CSI-RS), as well as the corresponding cell throughput.

\begin{table}[!htpb]
	\renewcommand\arraystretch{1.2}
	\centering
	\caption{Details of AI-empowered throughput optimization.}
	\label{tab:6}\vspace{2em}
	\begin{tabular}{|p{66pt}|l|}\hline
		AI-empowered\newline application & Intelligent throughput optimization\\\hline
		Optimized\newline objective&\tabincell{l}{Maximize the cell throughput}\\\hline
		\tabincell{l}{Input data}&\tabincell{l}{User location, BS azimuths, BS tilts,\\ BS powers, SSB RSRP}\\\hline
		\tabincell{l}{Output data}&\tabincell{l}{Antenna azimuths, antenna tilts \\ and transmit powers}\\\hline
		\tabincell{l}{AI algorithm}&\tabincell{l}{GPR and DNN}\\\hline
		\tabincell{l}{The role of AI}&\tabincell{l}{Regression and function fitting}\\\hline
	\end{tabular}
\end{table}

After the data collection, the data analysis and preprocessing are also necessary, where the feature selection and the data argumentation are carried out to meet the demands of the AI algorithm in terms of the accuracy, efficiency and robustness. On the other hand, inherited from conventional network optimization, the proposed approach is integrated with the scenario-specific communication model, which provides the performance simulation and theoretical analysis. The key of the proposed intelligent throughput optimization approach is to establish the mapping from network configuration parameters about the antenna azimuths, antenna tilts, and the transmit powers of the base station to the network performance metrics, i.e. the cell throughout, which is complex and difficult to be analyzed due to the massive connection and the complicated topology in 5G networks. Thus, the proposed approach leverages both the network state data and the scenario-specific communication model to fit the mapping relationship. Base on the established mapping, the antenna azimuths, antenna tilts, and the transmit powers of the base station is adjusted by the optimization methods and the output configuration parameters are deployed in the TTIN for the throughput performance evaluation. The proposed intelligent throughput optimization approach supports the advanced closed-loop control strategy, which is capable of adapting to the time-changed network environment.

\begin{figure}[!htpb]
\centering
\includegraphics[width=0.29\textwidth]{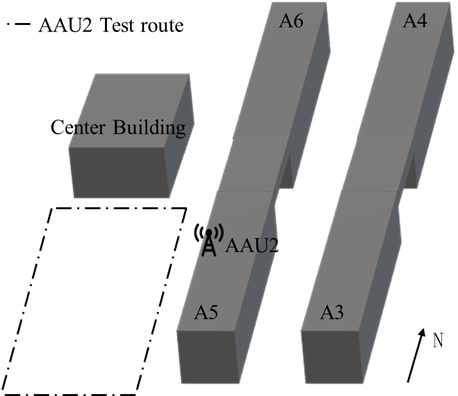}
\caption{Experiment scene of TTIN road test in CWV.}
\label{Fig.6}
\end{figure}

\begin{figure}[!htpb]
\centering
\includegraphics[width=0.4\textwidth]{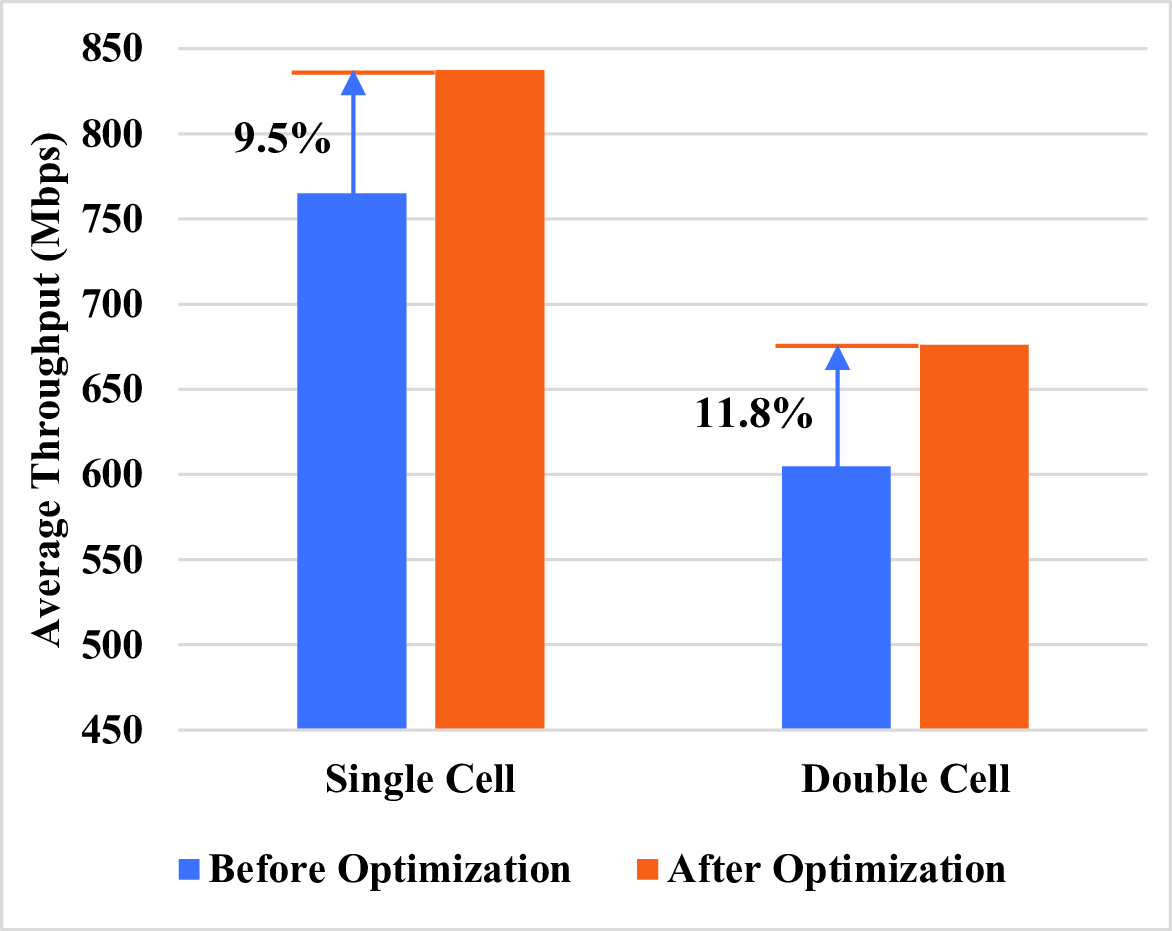}
\caption{Real-data experiment results of average throughput optimization.}
\label{Fig.7}
\end{figure}

Preliminary TTIN experiments show a remarkable performance gain in the intelligent throughput optimization. In the following, we introduce an experiment that was carried out in a real-world scenario in the CWV. The real experiment scene is depicted in Fig.~\ref{Fig.6}. For single cell case, AAU2 is selected and the test route is shown by the solid line while for double cell case, AAU1 and AAU2 are selected and the test route is displayed by the dotted line. The base station is equipped with $12\times8$ planar antenna array and adopts OFDM scheme in 5G NR Band n78 ($3.5\;{\rm{GHz}}$--$3.6\;{\rm{GHz}}$) with $100\;{\rm{MHz}}$ total bandwidth. The road test equipment adopts Pilot Matrix 5G automatic road test instruments to measure the synchronization signal block (SSB) beam pattern. Moreover, the Pilot Matrix is an automatic test solution for 5G NSA/SA wireless network air interface and service QoS/QoE, which is capable of testing and collecting data at the same time across the entire network and supporting the access of more test phones. In the real experiments, the road test equipment is connected to two mobile phone terminals, which are responsible for the uplink test and the downlink test. The mobile phones are Huawei Mate20 X. The performance of The RSRP data augmentation based on Gaussian process regression along the dimension of the user position and the BS antenna angle achieves a less-than $3\;{\rm{dBm}}$ mean absolute error. As shown in Fig.~\ref{Fig.7}, the subsequent deep neural network (DNN) based throughput optimization  can improve the average throughput approximately by $9.5\%$ for single cell case and $11.8\%$ for double cell case.

\subsection{Intelligent Massive MIMO}

\begin{figure}[!htpb]
\centering
\includegraphics[width=0.48\textwidth]{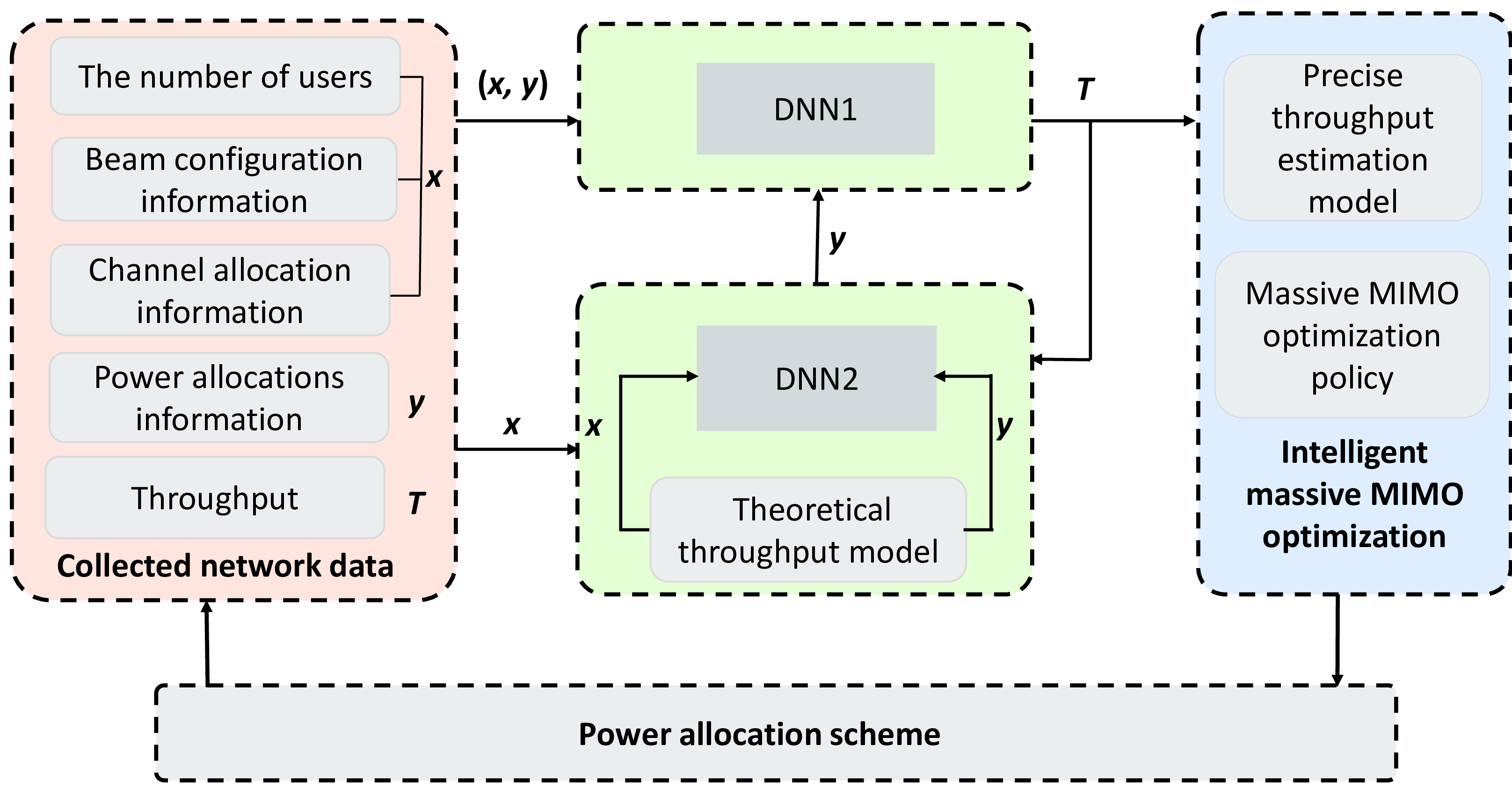}
\caption{Illustration of the intelligent massive MIMO optimization scheme.}
\label{Fig.8}
\end{figure}

In massive MIMO, the terminal antenna increases from 2R in the LTE to 4R/8R in the 5G, where the user throughput is accordingly improved to support the emerging bandwidth-thirsty applications including the UHD video, AR/VR and cloud game. To improve the throughput of the massive MIMO network, two DNNs called DNN1 and DNN2 are established to realize the precision throughput estimation as depicted in Fig.~\ref{Fig.8}. As listed in Table~\ref{tab:7}, DNN1 is used to estimate the throughput of the massive MIMO network through the collected network data, where the input features consist of the number of users, channel allocations, beam configurations and the power allocation, and the output of the DNN1 is the throughput of the actual system. DNN2 is utilized to provide (sub) optimal massive MIMO network optimization policy, where the input is the network state data in terms of the number of users, channel allocations, and beam configurations and the output is the power allocation scheme. Meanwhile, the theoretical throughput model with low precision is also built to provide the prior theory model knowledge and the sub-optical power allocation scheme.

\begin{table}[!htpb]
	\renewcommand\arraystretch{1.2}
	\centering
	\caption{Details of AI-empowered massive MIMO.}
	\label{tab:7}\vspace{2em}
	\begin{tabular}{|p{66pt}|l|}\hline
		AI-empowered application&Intelligent massive MIMO\\\hline
		Optimized objective&\tabincell{l}{Maximize total network throughput}\\\hline
		\tabincell{l}{Input data}&\tabincell{l}{The number of users, channel allocations,\\ beam  configurations, power allocation}\\\hline
		\tabincell{l}{Output data}&\tabincell{l}{Power allocation policy}\\\hline
		\tabincell{l}{AI algorithm}&\tabincell{l}{DNN}\\\hline
		\tabincell{l}{The role of AI}&\tabincell{l}{Regression and decision}\\\hline
	\end{tabular}
\end{table}

According to the estimated throughput from the well-trained DNN1, the DNN2 is fine-tuned through the stochastic gradient optimization algorithm to generate more appropriate power allocation scheme with better network performances. Integrated with two DNNs and the theoretical throughput model, the intelligent MIMO optimization approach is available to emerge optimal massive MIMO network optimization policy. The principle of the proposed intelligent massive MIMO network capacity improvement method is illustrated in Fig.~\ref{Fig.8}. In the proposed method, the throughput estimation model, i.e. DNN1, with strong generalization ability is trained with the data collected from the TTIN. Further, DNN 2 is pre-trained based on the low-precision theoretical throughput model to provide a sub-optimal policy. Finally, the DNN2 is modified and boosted according to the throughput estimation model to achieve the performance improvement. The proposed method is capable of capturing the characteristics of specific network scenarios through practical network data and the theoretical throughput model to achieve the intelligent massive MIMO network optimization.

\subsection{Intelligent Interference Coordination}

In the traditional interference coordination solutions, the time-varying user location information should be exchanged frequently and the interference coordination is generally implemented in the network elements. Thus, there is a large amount of data needed to be exchanged, which causes high computation loads and extra delay. To realize the real-time interference coordination and decrease the beam collision probability, long-term statistical data in terms of the user location and traffic hot spots is crucial to be fully utilized to establish the statistical beam collision probability model, which is further used to optimize the broadcast beam pattern and manage the frequency point intelligently.

\begin{figure}[!htpb]
\centering
\includegraphics[width=0.48\textwidth]{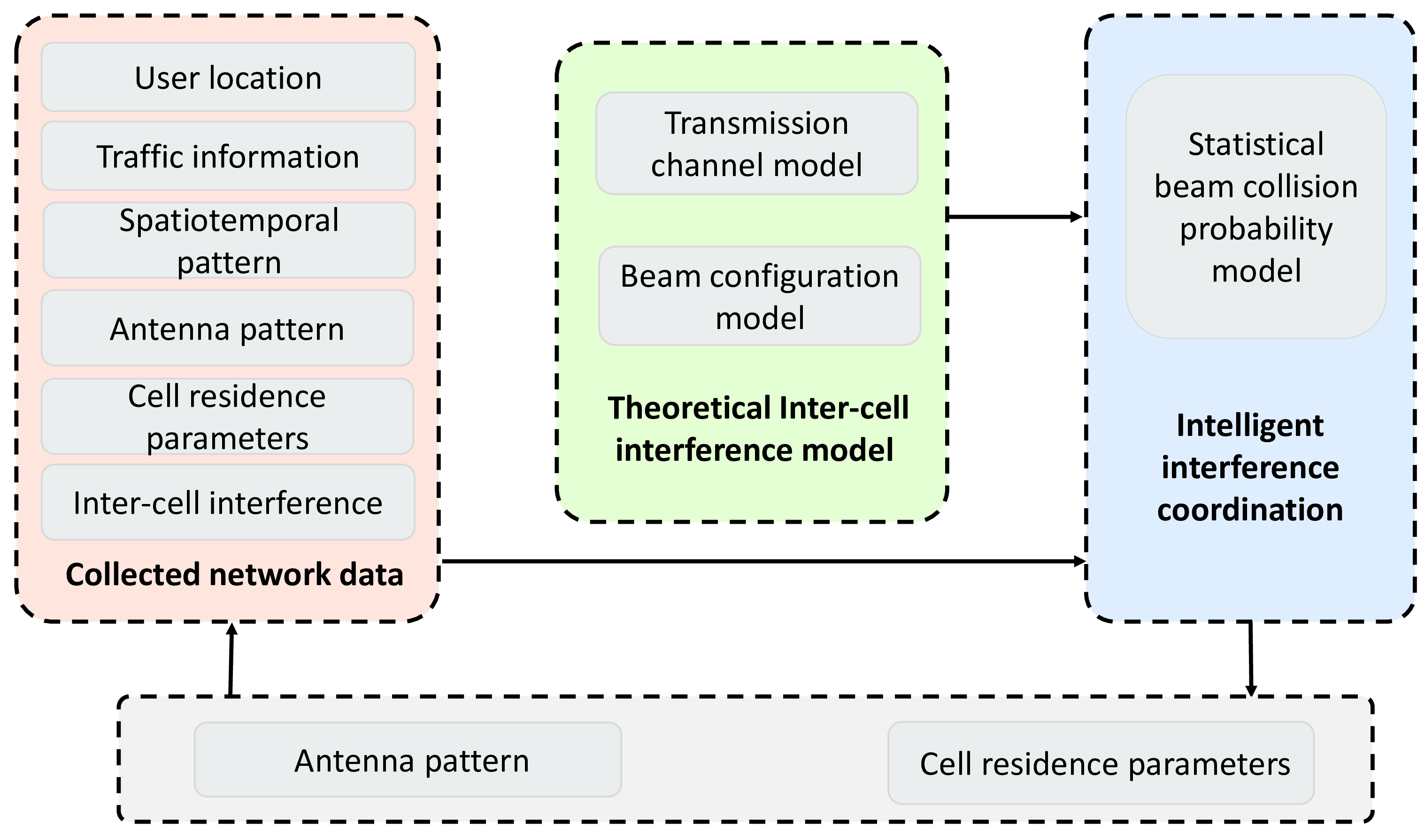}
\caption{Illustration of intelligent interference coordination approach.}
\label{Fig.9}
\end{figure}

Here, an intelligent beam collision avoidance and interference coordination model based on the multi-agent reinforcement learning is developed to select appropriate antenna pattern and cell residence parameters for decreasing the beam collision probability and improving the long-term network performances with the constraint of the QoS. In the proposed model as shown in Fig.~\ref{Fig.9}, a model-data-driven approach and multi-agent distributed scheme are utilized to adapt with the unstable network environment and decrease the number of exchanged messages.

As shown in Table~\ref{tab:8}, multi-dimensional network data is gathered, including comprehensive data from the users, base stations and inter-cell interference. On the user side, user location, the number of users, RSSI, transmission data of users, packet/frame loss number, associate starting time and ending time are collected comprehensively. On the base station side, the upload rate, download rate, number of server users, channel utilization, co-channel interference, neighborhood interference and air interface packets transceiving information are gathered. Secondly, the raw inter-cell interference model combined with the channel model under high loads and the beam configuration model related to the antenna pattern, the antenna orientation and the number of antennas is established. Finally, on the basis of the collected comprehensive network data and the raw inter-cell interference model, the adaptive statistical beam collision probability model using the multi-agent deep reinforcement learning is built and updated continuously through the real-time interaction between the agent and the network environment. In the proposed model, the environment information about the user location, traffic hot spots and spatiotemporal pattern, the action information about the antenna pattern and cell residence parameters and the reward information in terms of the corresponding interference from the inter-cell interference model are normalized as standard experience data for the training of the deep reinforcement learning agent through the trial and error mechanism. After the training phase of the deep reinforcement learning, the optimal antenna patterns and cell residence parameters are selected to obtain the minimum beam collision probability to realize the intelligent beam collision avoidance and interference coordination.

\begin{table}[!htpb]
	\renewcommand\arraystretch{1.2}
	\centering
	\caption{Details of AI-empowered interference coordination.}
	\label{tab:8}\vspace{2em}
	\begin{tabular}{|p{63pt}|l|}\hline
		AI-empowered application&Intelligent interference coordination\\\hline
		Optimized\newline objective&\tabincell{l}{Minimize the interference ratio}\\\hline
		\tabincell{l}{Input data}&\tabincell{l}{User location, the number of users,\\ RSSI, packet/frame loss number,\\ associate starting time and ending time,\\the upload rate, download rate, \\number of server users,\\ channel utilization,\\ co-channel interference,\\ neighborhood interference\\ and air interface, \\packets transceiving information}\\\hline
		\tabincell{l}{Output data}&\tabincell{l}{Antenna patterns and \\cell residence parameters}\\\hline
		\tabincell{l}{AI algorithm}&\tabincell{l}{DQN}\\\hline
		\tabincell{l}{The role of AI}&\tabincell{l}{Regression and decision}\\\hline
	\end{tabular}
\end{table}

\subsection{Intelligent Energy Conservation}

In the 5G/B5G networks, the power consumption of 5G is estimated to be $3$--$5$ fold higher than those of 4G. Therefore, how to reduce the electricity cost in the OPEX for operators has become one of the essential problems need to be addressed. As the power-consumption equipment, the energy cost of the AAU accounts for $80\%$ power consumption of those of the equipment in the 5G base station. Thus, it is crucial to decrease the power consumption of the AAU. There are several intelligent energy conservation innovations have been proposed, including the enhanced symbol shutdown (ESS), the enhanced channel shutdown (ECS) and the carrier wave shutdown (CWS).

In the ESS, the symbol shutdown is developed to decrease energy consumption by turning off the corresponding amplifier module through non-pilot, physical downlink control channel (PDCCH), secondary synchronization signal (SSS), physical broadcast signal (PBS) of the  downlink subframe. The enhanced method indicates that more time slots can be unoccupied, which further decreases the energy consumption for symbol shutdown. In the ECS, the resource block utilization ratios (RBUR) in the uplink and the downlink are monitored. Once the utilization ratio is less than the certain threshold, the channel shutdown will be triggered. The enhanced channel shutdown is specifically designed for massive MIMO, where the scale of the channel shutdown can be adjusted flexibly. In the CWS, the number of the utilized carrier is time-varying, where the capacity concurrent carrier can be turned off to decrease the energy consumption under the case of the low network load.

\begin{figure}[!htpb]
\centering
\includegraphics[width=0.48\textwidth]{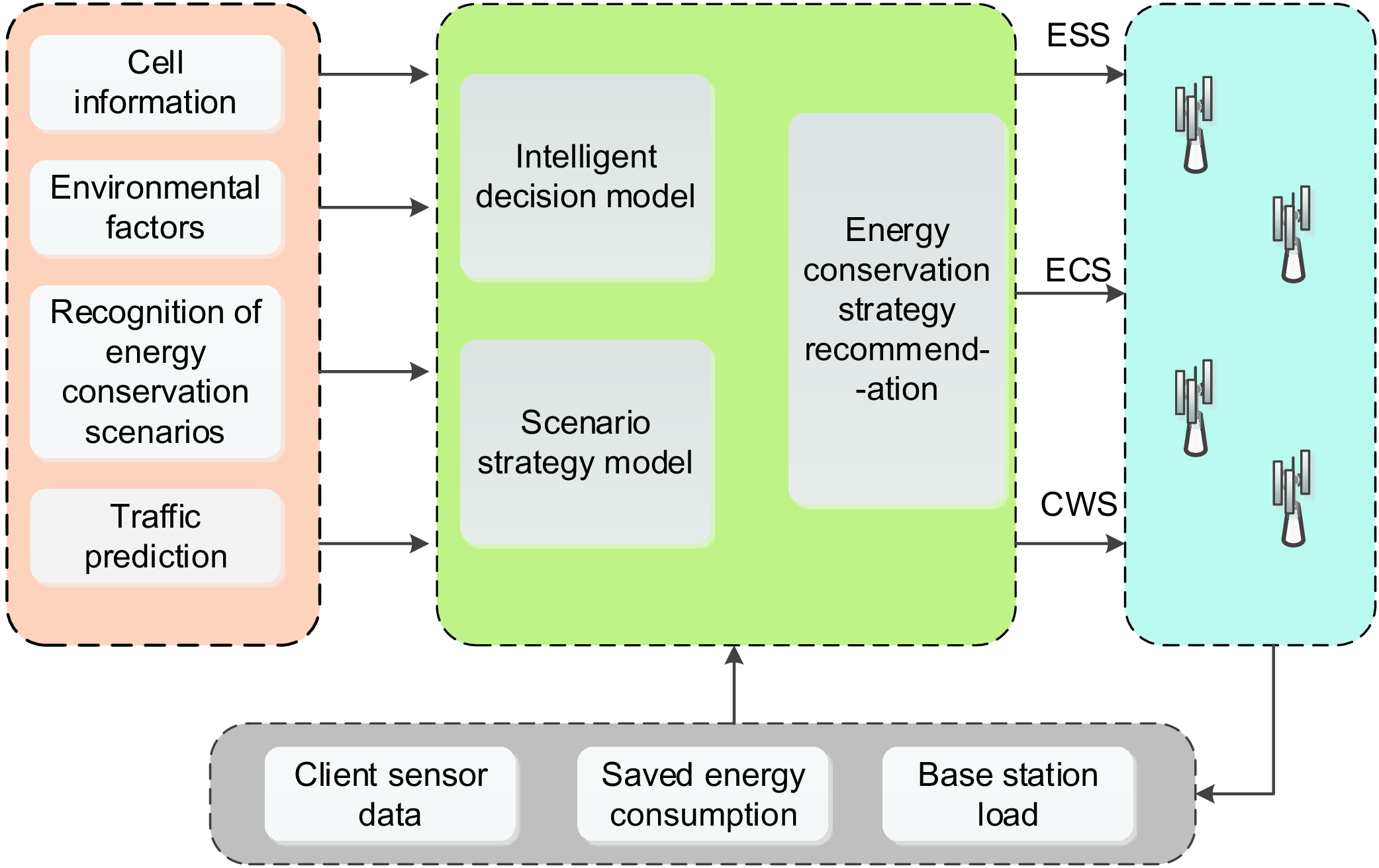}
\caption{Decision model of energy conservation strategy.}
\label{Fig.10}
\end{figure}

\begin{table}[!htpb]
	\renewcommand\arraystretch{1.2}
	\centering
	\caption{Details of AI-empowered energy conservation.}
	\label{tab:9}\vspace{2em}
	\begin{tabular}{|p{76pt}|l|}\hline
		AI-empowered\newline application &Intelligent energy conservation\\\hline
		Optimized objective&\tabincell{l}{Minimize the energy consumption}\\\hline
		\tabincell{l}{Input data}&\tabincell{l}{Performance data of base station, \\ MR/CDT data, Cell KPI}\\\hline
		\tabincell{l}{Output data}&\tabincell{l}{Energy conservation strategy \\recommendation}\\\hline
		\tabincell{l}{AI algorithm}&\tabincell{l}{ARIMA model, prophet model,\\ DNN, support vector machines}\\\hline
		\tabincell{l}{The role of AI}&\tabincell{l}{Regression, classification, decision}\\\hline
	\end{tabular}
\end{table}

As illustrated in Fig.~\ref{Fig.10}, the cell information, environment factors, recognition of energy conservation scenarios, and traffic prediction results are taken as the input of the model. The feature details of the AI technique are provided in Table~\ref{tab:9}. Through the intelligent decision model and the scenario strategy model, the best recommended energy conservation strategy is obtained and applied to the 5G base station. Moreover, we take the client sensor data, saved energy consumption and the 5G base station load as negative feedback terms to further optimize the model. In the TTIN, combined with advanced AI algorithms and wireless data platform, the energy conservation scenario can be recognized intelligently through massive historical data, where different energy conservation strategies are automatically recommended according to the real-time network loads.

\section{Conclusion}

The launch of TTIN marks a milestone in the development of 5G/B5G intelligent networks, for it overcomes the mutual isolation of on-site networks, big data, and AI techniques in the current 5G/B5G research. The open interfaces and the closed-loop control in this platform enable comprehensive true network data collection, standard dataset production, and intelligent data analysis, which facilitates \emph{in situ} inspection of AI algorithms and in turn improves the self-learning, self-optimizing, and self-managing capabilities of the networks. Additionally, the established experimental platform is open-source and ever-evolving, where open architecture and white-box hardware are utilized to provide extensible ecosystems. We argue that the B5G/6G paradigm is still in its infancy and there is a large spectrum of opportunities for the research community to develop new architectures, systems and applications, and to evaluate trade-offs in developing technologies for its successful deployment. Research institutes and vendors from all around the world are welcomed to verify and test diverse innovative solutions and methodologies on the TTIN for promoting the development of B5G/6G intelligent communication networks.

\bigskip
\bibliographystyle{IEEEtran}
\balance

\end{document}